\documentclass[11pt]{article}
\usepackage{pgf,tikz}
\usetikzlibrary{calc,arrows}
\usepackage[all]{xy}
\usepackage{amsmath}
\usepackage{amsfonts}
\usepackage{amssymb}
\usepackage{latexsym}
\usepackage{graphicx}
\usepackage{color}
\usepackage{amscd}
\usepackage{epsfig}
\usepackage{cite}
\usepackage{circuitikz}
\usepackage[perpage,symbol]{footmisc}
\usepackage{listings}
\usepackage{array}
\newtheorem{problem}{Problem}[section]

\newtheorem{lemma}[problem]{Lemma}
\newtheorem{theorem}[problem]{Theorem}

\title{Proof of spending in a block-chain system}
\author{Chunlei Liu\footnote{Shanghai Dengbi Comm. Tech. Co. Ltd., Shanghai, China. 714232747@qq.com}}
\date{}
\begin{document}
\maketitle
\abstract{We introduce proof of spending in a block-chain system. In this system the probability for a node to create a legal block is proportional to the total amount of coins it has spent in history.}

\section{\small{INTRODUCTION}}
\paragraph{}
In 2009, Satoshi Nakamoto \cite{Na} introduced the notion of block-chain as well as the notion of proof of work into P2P cash systems, giving birth to the famous Bitcoin, which is the first P2P cash implemented  in practise.

\paragraph{}
A cash system is a system in which nodes transfer coins to each other. A P2P cash system is a cash system in which transactions as well as datum built on transactions are broadcast to all nodes. A transaction is a collection of the following components: time of the blocktransaction, address of the payee, amount of payment, transaction fees,
unspent transactions, the change, and signature of the payer.

  
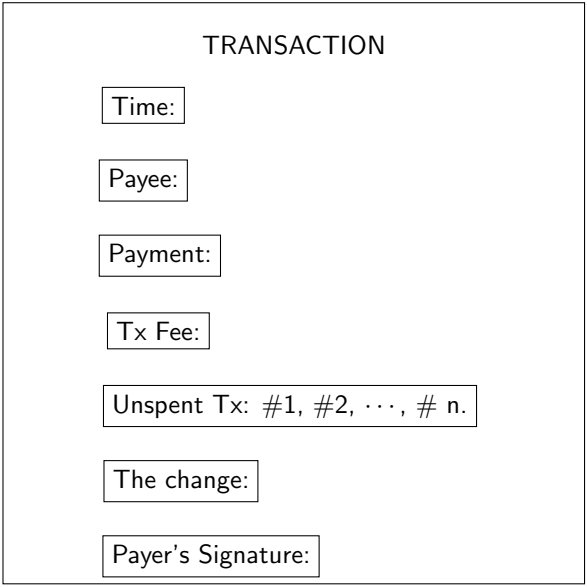
\begin{center}

\begin{tikzpicture}[font=\sffamily,>=triangle 45]
\draw (0,-2.5) node [shape=rectangle,scale=33,draw](){};
\node at (0,0.8) {TRANSACTION};
\draw (-2,0) node [shape=rectangle,draw] (Time) {Time:};
    \draw (-2,-1) node [shape=rectangle,draw] (Payee) {Payee:};
    \draw (-1.78,-2) node [shape=rectangle,draw] (Payment) {Payment:};
    \draw (-1.8,-3) node [shape=rectangle,draw] (Fees) {Tx Fee:};
    \draw (-0.05,-4) node [shape=rectangle,draw] (Unspent) {Unspent Tx: \#1, \#2, $\cdots$, \# n.};
    \draw (-1.5,-5) node [shape=rectangle,draw] (Change) {The change:};
    \draw (-1.1,-6) node [shape=rectangle,draw] (Payer) {Payer's Signature:};
  \end{tikzpicture}
\end{center}
\paragraph{}A block-chain system is a P2P cash system where transactions are collected into blocks, where blocks are chained one after another, and where only the longest block-chain is considered to be the correct one.
\paragraph{}A block in a block-chain system is a collection of the following components: time of the block, hash of the previous block, new transactions added to the block, address of the block creator, nonce such that the hash of the block begins with a number of zero bits.
  
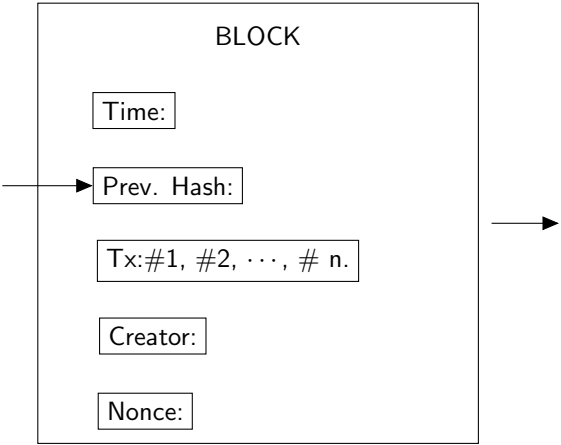
\begin{center}

\begin{tikzpicture}[font=\sffamily,>=triangle 45]
\draw (-1,-1.5) node [shape=rectangle,scale=25,draw](big){};
\node at (-1,1) {BLOCK};
\draw (-2.65,0) node [shape=rectangle,draw] (Time) {Time:};
    \draw (-1.4,-2) node [shape=rectangle,draw] (Transaction) {Tx:\#1, \#2, $\cdots$, \# n.};
    \draw (-2.2,-1) node [shape=rectangle,draw] (Prevhash) {Prev. Hash:};
    \draw (-2.5,-4) node [shape=rectangle,draw] (Nonce) {Nonce:};
    \draw (-2.4,-3) node [shape=rectangle,draw](Creator) {Creator:};
    \draw[->] (-4.4,-1)--(Prevhash);
    \draw[->] (big)--+(4,0);
  \end{tikzpicture}
\end{center}
\paragraph{}Let $D$ be a natural number. A proof of work system with target difficulty $D$ is a block-chain system where block $B$ must satisfy satisfies the threshold:
$${\rm nlz}({\rm hash}(B))\geq D.$$
Here ${\rm nlz}({\rm hash}(B))$ denotes the number of leading zero bits of ${\rm hash}(B)$.
\paragraph{}The expected time for a CPU to find a POW block is
$\approx2^D.$ And the expected time for $r$ CPUs to find a POW block is $\approx\frac{2^D}{r}.$  Therefore it is very difficult for the adversary to build the longest block-chain unless he has more CPUs than the honest party.
\paragraph{}In 2011, the notion of proof of stake was posted in bitcoin forum by a user named Quantunmechanic. Various proof of stake systems were then formulated, see, e.g. \cite{KN, BGM, NXT, Mi, BPS, DGKR, KRDO}.

\paragraph{}The simplest proof of stake system is the proof of balance system. A proof of balance system with target difficulty $D$ is a block-chain system where block $B$ chained after block-chain $C$ by node $A$ must satisfy the threshold:
$${\rm nlz}({\rm hash}(B))\geq D-\log_2(1+{\rm bal}(A;C)),$$
 where ${\rm bal}(A;C)$ is the balance of  node $A$ in block-chain $C$.

\paragraph{}The expected time for a party to find  a POB block chained to block-chain $C$ is $\approx\frac{2^D}{{\rm Bal}},$ where ${\rm Bal}$ is the balance of the party in block-chain $C$.
Suppose that the party transfers no coins to nodes outside the party, and assume that the transaction fees paid by the party in every block is a constant, say ${\rm FPB}$. Then the expected time for the party to build a long POB block-chain of length $L$ is
$\approx\frac{2^D\log L}{{\rm Rwd}-{\rm FPB}},$
where ${\rm Rwd}$ is the coins rewarded to a block creator. Thus, the expected time for a party without spending to build a long POB block-chain of length $L$ is
$\approx\frac{2^D\log L}{{\rm Rwd}}.$
It follows that, the adversary who never spends his coins can build  a long block-chain secretely which in a long run, would outpace the block-chain maintained by the honest party. The same philosophy can be applied to attack other kinds of proof of stake systems, see, e.g. \cite{Bu, Po}.
\paragraph{}
In this paper we present a proof of spending system. In this system the expected time for a node to create a block is inverse proportional to the total amount of the coins it has spent in history. We shall see  that, in the proof of spending system, the adversary trying to build a longest block-chain would earn nothing.

\section{PROOF OF SPENDING}

\paragraph{}We now present a proof of spending system.

\paragraph{}The proof of spending system with target difficulty $D$ is a block-chain system where block $B$ chained after block-chain $C$ by node $A$ must satisfy the threshold:
$${\rm nlz}({\rm hash}(B))\geq D-\log_2(1+{\rm spn}(A;C)),$$
where ${\rm spn}(A,C)$ is the total amount of coins spent by node $A$ in block-chain $C$.
\paragraph{}We call a block in a proof of spending system a PSP block. One can prove the following.
\begin{lemma}The expected time for a party to find  a PSP block chained after block-chain $C$ is
$\approx\frac{2^D}{{\rm Spn}},$
where ${\rm Spn}$ is the amount of coins spent by the party in block-chain $C$.\end{lemma}
\paragraph{}We now prove the following.
\begin{lemma}Suppose that a party is going to build a long PSP block-chain, and assume that no nodes outside the party would transfer coins to the party. Then the coins spent per block by the party is
$\leq\frac{{\rm Rwd}}{{\rm FPC}},$ where ${\rm FPC}$ is the amount of transaction fees per coin.\end{lemma}
{\it Proof.}  Suppose that the contrary is true. Let ${\rm SPB}$ be the amount of coins spent per block by the party Then whenever the party produces a PSP block, the balance of the party decreases at least by ${\rm SPB}\times{\rm FPC}-{\rm Rwd}>0$. This would forbid the party to build a long block-chain, and thus contradicts to the assumption of the lemma. The lemma is proved.
\paragraph{}We now prove the following.
\begin{theorem}Suppose that the coins spent by a party in every block is $\frac{{\rm Rwd}}{{\rm FPC}}\cdot{\rm WR}$, where ${\rm WR}\leq1$ is a positive constant. Then the expected time for the party to build a long PSP block-chain of length $L$ is
$$\approx \frac{2^D}{\rm WR}\cdot\frac{\rm FPC}{{\rm Rwd}}\cdot(\gamma+\log L),$$
where $\gamma$ is the Euler constant.\end{theorem}
{\it Proof.}  Since the coins spent by the party in the first $i$ blocks of the block-chain is $i\cdot \frac{{\rm Rwd}}{{\rm FPC}}\cdot{\rm WR}$, the expected time for the party to produce the $(i+1)$-th block is
$\approx\frac{2^D}{\rm WR}\cdot\frac{\rm FPC}{{\rm Rwd}}\cdot\frac{1}{i}.$ So the expected time for the party to build a long block-chain of length $L$  is
$$\approx\frac{2^D}{\rm WR}\cdot\frac{\rm FPC}{{\rm Rwd}}\sum_{i=1}^L\frac{1}{i}$$
$$\approx \frac{2^D}{\rm WR}\cdot\frac{\rm FPC}{{\rm Rwd}}\cdot(\gamma+\log L).$$ The theorem is proved.
\paragraph{}Note that at the growing stages of the network, a proof of spending system is nearly a proof of work system, and hence is secure. After the network is grown up, the coins spent by the honest party in every block is $\approx\frac{{\rm Rwd}}{{\rm FPC}}$.
Suppose that the adversary  wants to  build a long PSP block-chain in shortest time. His best strategy is to  transfer $\frac{{\rm Rwd}}{{\rm FPC}}\cdot{\rm WR}$ coins to himself in every block with ${\rm WR}<1$. Therefore it is difficult for the adversary to built the longest block-chain alone.

\section{PROOF OF RECENT SPENDING}

\paragraph{}We now present a proof of recent spending system.

\paragraph{}Let $F$ be a natural number. The proof of recent spending system with target difficulty $D$ and freshness $F$ is a block-chain system where block $B$ chained after block-chain $C$ by node $A$ must satisfy the threshold:
$${\rm nlz}({\rm hash}(B))\geq D-\log_2(1+{\rm spn}(A;C^{[F]})),$$
where $C^{[F]}$ is the last segment of $C$ of length $F$, and ${\rm spn}(A,C^{[F]})$ is the total amount of coins spent by node $A$ in the chain segment $C^{[F]}$.
\paragraph{}We call a block in a proof of recent spending system a PRS block.
As in the last section, we can prove the following two lemmas.
\begin{lemma}The expected time for a party to find  a PRS block chained after block-chain $C$ is
$\approx\frac{2^D}{{\rm RS}},$
where ${\rm RS}$ is the amount of coins spent by the party in chain segment $C^{[F]}$.\end{lemma}
\begin{lemma}Suppose that a party is going to build a long PRS block-chain, and assume that no nodes outside the party would transfer coins to the party. Then the coins spent per block by the party is
$\leq\frac{{\rm Rwd}}{{\rm FPC}},$ where ${\rm FPC}$ is the amount of transaction fees per coin.\end{lemma}
We now prove the following.
\begin{theorem}Suppose that the coins spent by a party in every block is $\frac{{\rm Rwd}}{{\rm FPC}}\cdot{\rm WR}$, where ${\rm WR}\leq1$ is a positive constant. Then the expected time for the party to build a long PRS block-chain of length $L$ is
$$\approx \frac{2^D}{F}\cdot\frac{\rm FPC}{{\rm Rwd}}\cdot\frac{1}{\rm WR}\cdot L.$$\end{theorem}
{\it Proof.}  We have, when $C$ is long, $${\rm spn}(A,C^{[E]})=F\cdot\frac{{\rm Rwd}}{{\rm FPC}}\cdot{\rm WR}.$$
 So, when $i$ is large, the expected time for the party to produce the $i$-th block is
$$\approx\frac{2^D}{F}\cdot\frac{\rm FPC}{{\rm Rwd}}\cdot\frac{1}{\rm WR}.$$ So the expected time for the party to build a long block-chain of length $L$  is
$$\approx \frac{2^D}{F}\cdot\frac{\rm FPC}{{\rm Rwd}}\cdot\frac{1}{\rm WR}\cdot L.$$The theorem is proved.
\paragraph{}Suppose that the adversary  wants to  build a long PRS block-chain alone in shortest time. His best strategy is to  transfer $\frac{{\rm Rwd}}{{\rm FPC}}\cdot{\rm WR}$ coins to himself in every block with ${\rm WR}$ close to 1. The transaction fees he must pay in very block is ${\rm Rwd}\cdot{\rm WR}$. So he earns ${\rm Rwd}\cdot(1-{\rm WR})$ per block. As the time for him to create a block is $$\approx\frac{2^D}{F}\cdot\frac{\rm FPC}{{\rm Rwd}}\cdot\frac{1}{\rm WR},$$
his earning, per unit time, is
$$\approx\frac{F}{2^D}\cdot\frac{{\rm Rwd}^2}{{\rm FPC}}\cdot{\rm WR}(1-{\rm WR}),$$
which is very small. It follows that the proof of recent spending system is secure, and is very secure if $$\frac{F}{2^D}\cdot\frac{{\rm Rwd}^2}{{\rm FPC}}$$
is small.

\section{SPENDING OF OLD COINS}

\paragraph{}We now present a proof of spending of old coins system.
\paragraph{}We begin with the definition of coin age. The age of coin in a block chain is defined to be the length from the last transaction of the coin to the end of the block chain. The age of coin $co$ in block-chain $C$ is denoted as ${\rm age}(co;C)$.
\paragraph{}Let $E$ be a natural number. The proof of spending of old coins system with target difficulty $D$ and experience $E$ is a block-chain system where block $B$ chained after block-chain $C$ by node $A$ must satisfy the threshold:
$${\rm nlz}({\rm hash}(B))\geq D-\log_2(1+{\rm spn}(A;C,E)),$$
where ${\rm spn}(A,C,E)$ is the total amount of coins of age at least $E$ spent by node $A$ in the block-chain $C$.
\paragraph{}We call a block in a proof of spending of old coins system a PSO block.
As in the last section, we can prove the following two lemmas.
\begin{lemma}The expected time for a party to find  a PSO block chained after block-chain $C$ is
$\approx\frac{2^D}{{\rm SO}},$
where ${\rm SO}$ is the amount of coins of age at least $E$ spent by the party in block-chain $C$.\end{lemma}
\begin{lemma}Suppose that a party is going to build a long PSO block-chain, and assume that no nodes outside the party would transfer coins to the party. Then the coins spent per block by the party is
$\leq\frac{{\rm Rwd}}{{\rm FPC}},$ where ${\rm FPC}$ is the amount of transaction fees per coin.\end{lemma}
We now prove the following.
\begin{theorem}Suppose that the coins spent by a party in every block is $\frac{{\rm Rwd}}{{\rm FPC}}\cdot{\rm WR}$, where ${\rm WR}\leq1$ is a positive constant. Then the expected time for the party to build a long PSO block-chain of length $L$ is
$$\approx 2^D\cdot\frac{\rm FPC}{{\rm Rwd}}\cdot\frac{1}{\rm WR}\cdot \log L.$$\end{theorem}
{\it Proof.}  We have, when $C$ is long, $${\rm spn}(A;C,E)\approx{\rm len}(C)\cdot\frac{{\rm Rwd}}{{\rm FPC}}\cdot{\rm WR},$$ where ${\rm len}(C)$ is the length of $C$.
 So the expected time for the party to build a block-chain of length $L$  is
$$\approx \sum_{i=1}^L\frac{2^D}{i}\cdot\frac{\rm FPC}{{\rm Rwd}}\cdot\frac{1}{\rm WR}$$
$$\approx 2^D\cdot\frac{\rm FPC}{{\rm Rwd}}\cdot\frac{1}{\rm WR}\cdot \log L.$$The theorem is proved.
\paragraph{}Suppose that the adversary  wants to  build a long PRS block-chain alone in shortest time. His best strategy is to  transfer $\frac{{\rm Rwd}}{{\rm FPC}}\cdot{\rm WR}$ coins to himself in every block with ${\rm WR}$ close to 1. His balance must be greater than $E\cdot \frac{{\rm Rwd}}{{\rm FPC}}\cdot{\rm WR}$. If $E\geq \frac{{\rm FPC}}{{\rm Rwd}}\cdot{\rm CN}$, where ${\rm CN}$ is the coins of the network, then the balance of the adversary must be greater than ${\rm CN}\cdot{\rm WR}$. It follows that the proof of spending of old coins is secure if $E$ is large.
\section{RECENT SPENDING OF OLD COINS}

\paragraph{}We now present a proof of recent spending of old coins system.
\paragraph{}The proof of recent spending of old coins system with target difficulty $D$, freshness $F$ and experience $E$ is a block-chain system where block $B$ chained after block-chain $C$ by node $A$ must satisfy the threshold:
$${\rm nlz}({\rm hash}(B))\geq D-\log_2(1+{\rm spn}(A;C^{[F]},E)),$$
where ${\rm spn}(A,C^{[F]},E)$ is the total amount of coins of age at least $E$ spent by node $A$ in the segment $C^{[F]}$.
\paragraph{}We call a block in a proof of spending of old coins system a RSO block.
As in the last section, we can prove the following two lemmas.
\begin{lemma}The expected time for a party to find  a RSO block chained after block-chain $C$ is
$\approx\frac{2^D}{{\rm SO}},$
where ${\rm SO}$ is the amount of coins of age at least $E$ spent by the party in segment $C^{[F]}$.\end{lemma}
\begin{lemma}Suppose that a party is going to build a long RSO block-chain, and assume that no nodes outside the party would transfer coins to the party. Then the coins spent per block by the party is
$\leq\frac{{\rm Rwd}}{{\rm FPC}},$ where ${\rm FPC}$ is the amount of transaction fees per coin.\end{lemma}
We now prove the following.
\begin{theorem}Suppose that the coins spent by a party in every block is $\frac{{\rm Rwd}}{{\rm FPC}}\cdot{\rm WR}$, where ${\rm WR}\leq1$ is a positive constant. Then the expected time for the party to build a long RSO block-chain of length $L$ is
$$\approx \frac{2^D}{F}\cdot\frac{\rm FPC}{{\rm Rwd}}\cdot\frac{1}{\rm WR}\cdot L.$$\end{theorem}
{\it Proof.}  We have, when $C$ is long, $${\rm spn}(A;C,E^{[F]})\approx F\cdot\frac{{\rm Rwd}}{{\rm FPC}}\cdot{\rm WR}.$$
 So the expected time for the party to build a block-chain of length $L$  is
$$\approx \frac{2^D}{F}\cdot\frac{\rm FPC}{{\rm Rwd}}\cdot\frac{1}{\rm WR}\cdot L.$$The theorem is proved.
\paragraph{}Suppose that the adversary  wants to  build a long PRS block-chain alone in shortest time. His best strategy is to  transfer $\frac{{\rm Rwd}}{{\rm FPC}}\cdot{\rm WR}$ coins to himself in every block with ${\rm WR}$ close to 1. His balance must be greater than $E\cdot \frac{{\rm Rwd}}{{\rm FPC}}\cdot{\rm WR}$. If $E\geq \frac{{\rm FPC}}{{\rm Rwd}}\cdot{\rm CN}$, where ${\rm CN}$ is the coins of the network, then the balance of the adversary must be greater than ${\rm CN}\cdot{\rm WR}$. The transaction fees he must pay in very block is ${\rm Rwd}\cdot{\rm WR}$. So he earns ${\rm Rwd}\cdot(1-{\rm WR})$ per block. As the time for him to create a block is $$\approx\frac{2^D}{F}\cdot\frac{\rm FPC}{{\rm Rwd}}\cdot\frac{1}{\rm WR},$$
his earning, per unit time, is
$$\approx\frac{F}{2^D}\cdot\frac{{\rm Rwd}^2}{{\rm FPC}}\cdot{\rm WR}(1-{\rm WR}),$$
which is very small. It follows that the proof of recent spending system is secure, and is very secure if $$\frac{F}{2^D}\cdot\frac{{\rm Rwd}^2}{{\rm FPC}}$$
is small. It follows that the proof of spending of old coins is secure if $E$ is large.
\section{{\small CONCLUSION}}
We have proposed two block-chain systems: proof of spending and proof of recent spending. The proof of spending system is more efficient, and the proof of recent spending system is more secure.

\end{document}